\documentclass{article}
\usepackage{spconf,amsmath,graphicx}
\usepackage{cite}
\usepackage{amsmath,amssymb,amsfonts}
\usepackage{nth}
\usepackage{graphicx}
\usepackage{textcomp}
\usepackage{xcolor}
\usepackage{booktabs}
\usepackage{multirow}
\usepackage{algorithm}
\usepackage{algpseudocode}
\usepackage{algorithmicx}
\usepackage{indentfirst}
\usepackage{subfigure}
\title{FILTER AND EVOLVE: PROGRESSIVE PSEUDO LABEL REFINING FOR SEMI-SUPERVISED AUTOMATIC SPEECH RECOGNITION}
\name {Zezhong~Jin$^{1}$, Dading Zhong$^{2}$, Xiao Song$^{2}$, Zhaoyi Liu$^{3}$, Naipeng Ye$^{4}$, Qingcheng Zeng$^{5}$}
\address{ $^1$Dept. of Electronic and Information Engineering\\
 The Hong Kong Polytechnic University, Hong Kong SAR, China\\
 $^2$School of Electronic and Computer Engineering, Peking University, China\\
 $^3$imec-Distrinet, Computer Science, KU Leuven, Leuven, Belgium\\
 $^4$Anhui Provincial Key Laboratory of Multimodal Cognitive Computation\\
 School of Computer Science and Technology, Anhui University, Hefei, China\\
 $^5$Department of Linguistics, Northwestern University, USA\\}
\begin{document}
%
\maketitle
\begin{abstract}
Fine-tuning self-supervised pre-trained models using pseudo-labels can effectively improve speech recognition performance. But, low-quality pseudo-labels can misguide decision boundaries and degrade performance. We propose a simple yet effective strategy to filter low-quality pseudo-labels to alleviate this problem. Specifically, pseudo-labels are produced over the entire training set and filtered via average probability scores calculated from the model’s output. Subsequently, an optimal percentage of utterances with high probability scores are considered reliable training data with trustworthy labels. The model is iteratively updated to correct
the unreliable pseudo-labels to minimize the effect of noisy labels. The process above is repeated until unreliable pseudo-labels have been adequately corrected. Extensive experiments
on LibriSpeech show that these filtered samples
enable the refined model to yield more correct predictions, leading to better ASR performances under various experimental
settings.
\end{abstract}
\begin{keywords}
self-supervised models, semi-supervised learning, ASR, pseudo-labeling
\end{keywords}
\section{Introduction}
Benefiting from the availability of large-scale annotated data and the evolution of optimization algorithms, more capable and more complex deep-learning models have been developed for automatic speech recognition (ASR) \cite{vaswani2017attention,chan2016listen}. Nonetheless, curating large-scale transcribed speech data is time-consuming, especially for low-resource languages.

Recently, numerous research efforts have been dedicated towards reducing the over-dependency on annotated speech data. These efforts including data augmentation \cite{park2019specaugment,ragni2014data}, local prior matching \cite{hsu2020semi}, representation learning \cite{baevski2020wav2vec,hsu2021hubert,kahn2020libri}, pseudo-labeling (PL) \cite{kahn2020self,xu2020iterative,higuchi2021momentum}, and consistency regularization \cite{qi2022improved,chen2021semi}. PL, as an important method in semi-supervised learning, trains the teacher model with labeled data and uses the prior knowledge of the teacher model to generate pseudo-labels, followed by utilizing pseudo-labels and few labeled data to train a student model. PL-based method can achieve good performance with few labeled data, this will greatly save the consumption of manual labeling.
The author in \cite{xu2020iterative,higuchi2021momentum} show that PL can achieve extraordinary results on ASR tasks.

Despite the good performance, current PL-based methods still face great challenges. Firstly, many of pseudo-labels are incorrect due to the poor calibration of teacher models. For poorly calibrated networks, an incorrect prediction might have high conﬁdence, which can misguide the decision boundaries and degrade performance. However, a good teacher model does not necessarily make PL-based methods effective. The distracting pseudo-labels are another obstacle. Many recent studies have involved learning from noisy data. Under such a scenario, PL-based methods tend to underperform due to the generation of noisy pseudo-labels. Choosing high-confidence pseudo-labels while reducing noise is a challenge that
needs to be solved. Additionally, many pseudo-labels are wasted due to mediocre filtering strategies and incorrect selection criteria. We require an ideal filtering threshold that can strike a good balance between the quality and quantity of pseudo-labels
to make the best use of pseudo-labels.

In this paper, we focus on the PL-based methods and aim at tackling these three challenges. (1) We propose a framework that generates a teacher model by fine-tuning self-supervised pre-trained model. Result show that this approach can significantly reduce the chance of poor network calibration. (2) To reduce noisy pseudo-labels, we choose pseudo-labels based on the average probability score of the model's output. With numerous trials using various label data amounts, we demonstrate the efficacy of this approach.(3) Based on the distribution of the filtered samples, we determine the best filtering threshold for sample selection.  

\section{PROPOSED METHOD}
This section describes how we fine-tune the pre-trained Wav2vec2.0 model \cite{baevski2020wav2vec}, followed by explaining our strategy (namely iterative pseudo-labeling) for selecting high quality pseudo-labels. Finally, a general strategy to obtain the filtering threshold for selecting pseudo labels is described. 
\subsection{Pre-training and Fine-tuning}
Wav2vec 2.0 \cite{baevski2020wav2vec} is the unsupervised pre-trained model published by Facebook in 2020. Its core idea is to construct a self supervised training target through Vector Quantization (VQ), make a lot of masks on the encoder output, and use the contrast learning loss function to train.

During fine-tuning, models are optimized by minimizing a connectionist temporal classification (CTC) loss. Let ${\cal X}$ $=\left\{\mathrm{x}_t \in \mathbb{R}^D; i=1,\dots,T\right\}$, where ${\mathrm{x}}_i$ denotes a $D$-dimensional vector at the $i$-$\rm{th}$ frame. Let ${\cal Y}$ $= \left\{ y_l \in {\cal V}; l=1,\dots,L \right\}$, where $y_l$ is a target token at position $l$, and ${\cal V}$ comprises 29 character tokens and 3 special symbols (start, end, and a word boundary token). 
CTC \cite{graves2006connectionist} is an objective function for end-to-end (E2E) sequence learning, which performs a frame-level alignment between the input sequence $\cal X$ and the output sequence $\cal Y$ by introducing a special \texttt{<blank>} token. With the alignments $\cal Z$ $=\{ z_t \in \mathcal{V} \cup \{\texttt{<blank>}\} ; t=1,\dots,T\}$, CTC models the conditional probability of ${\cal Y}$ given ${\cal X}$  by searching over all paths:
\begin{equation}
    \label{eq:p_ctc}
    P_{\mathrm{ctc}} ({\cal Y} | {\cal X}) = \sum_{{\cal Z} \in \beta ({\cal Y})} \prod_{t=1}^{T} P_{\mathrm{ctc}} (z_t | {\cal X}),
\end{equation}
where $\beta ({\cal Y})$ is the set of all paths (frame alignments) compatible with ${\cal Y}$. The Loss function of CTC is defined as:
\begin{equation}
    \label{eq:loss}
    L_{\mathrm{ctc}} =-\mathop{\sum}_{({\cal X},{\cal Y})\in S} {\rm ln}P_{\mathrm{ctc}} ({\cal Y} | {\cal X}),
\end{equation}
where $\cal S$ contains all training data.

\subsection{Purifying pseudo-labels}
\begin{algorithm}[!t]
  \caption{Iteration filtering and selection of pseudo-labels} 
  \label{alg::IPL}
  {\small
  \begin{algorithmic}[1]
    \Require
      $D_{lab},D_{unlab}$: labeled and unlabeled data
      
      $A$: an ASR model architecture
      
      $\theta$: model weight

      $iter_{max}$: Maximum number of iterations

      $epoch_{max}$: Maximum number of epochs
    \State Train a teacher model $M_{\theta_0}$ with architecture $A$ on $D_{lab}$ using Eq.~\ref{eq:p_ctc} and Eq.~\ref{eq:loss};
    \For{$t=1,\dots,iter_{max}$}
    \State Predict pseudo-labels for $D_{unlab}$ using $M_{\theta_{t-1}}$
    \State Filter the pseudo-labels to get $D_{pseudo}$ using Eq.~\ref{score} and Eq.~\ref{filtering}
    \State Fusing $D_{pseudo}$ with $D_{lab}$
\For{$epoch=1,\dots,epoch_{max}$}
    \For{${\cal S} \in D_{pseudo} \cup D_{lab}$}
    \State Obtain ${\cal X}\sim {\cal S}$;
    \State Obtain $P^{\theta_t}_{ctc}(Y|X)$ using Eq.~\ref{eq:p_ctc}
    \State Obtain ${\cal Y} \sim P^{\theta_t}_{ctc}({\cal Y}|{\cal X})$;
    \State Compute CTC loss using groundtruth labels and pseudo-labels
    \State Update $\theta_t$ based on the CTC loss using backpropagation.
    \EndFor
    \EndFor
    \EndFor
    \State \Return {$M_{\theta_{t}} (t=iter_{max})$}
  \end{algorithmic}
  }
\end{algorithm}
The method we used for fine-tuning is described in Algorithm \ref{alg::IPL}.

During the prediction, after input the audio to the model, the model use N-dimensional vector to represent the probability score for each frame, where $N$ is the number of target token. With $T$ frames, we can get a $C_{_{N \times F}}$ matrix. Since this project does not involve a language model, we can calculate the average probability score for each utterance as follows:
\begin{equation}
\label{score}
    score = \frac{1}{T}\sum_{i=1}^{T} {\rm argmax} 
     (C^{\mathrm {_{trans}}}_{_{N \times T}}),
\end{equation}
Predicted results with high score is considered as high confidence pseudo labels.

Let $\mathop{{\cal D}}\limits^\sim =\{( \mathop{y_i}\limits^\sim, s_i)\}_{i=1}^{N_{_U}}$ be a pseudo-label dataset with $N_{_U}$ samples, where $\mathop{y_i}\limits^{\sim}$ is a pseudo label generated by model prediction and $s_i$ is the corresponding score to $\mathop{y_i}\limits^{\sim}$. Let $\cal E$ denotes the decision boundary, we can filter the pseudo-labels as follows:
\begin{equation}
\label{filtering}
   \mathop{{\cal D^\prime}}\limits^\sim= \left\{  
\begin{aligned}
y_i &, & if s_i> {\cal E} \\
0  &, &else \\
\end{aligned}
    \right\}
\end{equation}
where $\mathop{{\cal D^\prime}}$ denotes the filtered pseudo-labeled set. We select pseudo-labels with a score above the decision boundary. Section 2.3 illustrates the strategy for optimizing the decision boundary.
\renewcommand{\algorithmicrequire}{\textbf{Input:}}
\renewcommand{\algorithmicensure}{\textbf{Output:}}
\subsection{Strategy for optimize the filtering threshold}
In order to make the most of pseudo-labels, we want to find an optimal decision boundary to get the ideal balance between quality and quantity of pseudo-labels. After conducting experiments to filter the pseudo-labels based on ground-truth label and score, we find an extensive overlap between samples filtered by these two protocol. Based on the score distribution of filtered samples, we can choose a score value that makes the number of filtered samples approximately equal to the overlapping samples. Inspired by this, we make an assumption that the optimal filtering threshold is nearly equal to the score value we select.

We suggested a plan of action to support this presumption. Let decision boundary ${\cal E}_u = u \times c$, where $u$ means the number of updates, after three iterations, we update the threshold. ${\cal E}_u$ denotes the value of boundary at $u$-$\rm{th}$ update and $c$ is a hyperparameter which indicates the step size of the threshold change for each update. As the number of update increases, when the model performance begins to decline, with $c$ determined, the optimal threshold is $E_{u-1}$. Under $E_{u-1}$, if the number of filtered data is approximately equal to the overlapped samples and $E_{u-1}$ is roughly equal to the value we select, the assumption will be approved.
\vspace{-1.0mm}
\section{EXPERIMENTS SETTING}
\subsection{Datasets}
The experiments are carried out using the LibriSpeech
(LS) \cite{panayotov2015librispeech} datasets. LS is a
corpus of English which consist of 960 hours training data (split into train-clean-100, train-clean-360 and train-other-500). Each audio file has a sample frequency of 16 kHz and takes for roughly 10 seconds.

LS is divided into two parts, labeled and unlabeled. For instance, LS-100/LS-860 is a combination of 860 hours of unlabeled data and 100 hours of labeled data. LS-100/LS-860, LS-10/LS-90, LS-10/LS-860, and LS-1/LS-90 are the various semi-supervised settings that we have set up.


\subsection{Training Configuration}
We choose the base Wav2vec2.0 pre-trained model from open source projects Fairseq \cite{wang2020fairseq}.
In the fine-tuning stage, We optimize with Adam and a tri-state \cite{zhu2020adaptive} rate schedule where the learning rate is warmed up for the first 10\% of updates, held constant for the next 40\% and then linearly decayed for the remainder. We set layer drop to 0.05 in LS-10 experiment and 0.1 in other experiments. Similarly, we also set initial learning rate $=0.0005$ for LS-10 and LS-1, and $0.0003$ for other experiments. The best model is obtained according to the best validation performance during training stage.
\subsection{Evaluation}
Following the protocols in \cite{baevski2020wav2vec}, we evaluate the prediction
results by WER metric on LS clean and other test datasets. The formula is defined as:
\begin{equation}
    WER=\frac{S+D+I}{N}=\frac{S+D+I}{S+D+H}
\end{equation}
where $S$ is the number of words replaced, $D$ is the number of words deleted, $I$ is the number of words inserted, and $H$ is the number of correct words.

\section{RESULTS AND DISCUSSION}
The experiment results of our method are compared to other existing methods in this section. Baseline indicates the model trained on the LS-labelled dataset in \cite{baevski2020wav2vec}, the results are shown in Table 1. To make a fair comparison with other methods without the interference of language models, all of our experiments do not include language models. In this case, we save training time and reduce the time needed for the model to predict pseudo-labels.
\begin{table}[b]
\small
\centering
\label{table::wav2vec_facebook}
\renewcommand{\arraystretch}{1}
    \begin{tabular}{cccc}
\toprule
Pre-train Model& Fine-tune & Test(clean)& Test(other)\\
\midrule
\multirow{3}{*}{Wav2Vec2.0 base}&1 hours &24.50 &29.70 \\
& 10 hours&11.10 &17.60\\
&100 hours &6.10 &13.30\\
\bottomrule
\end{tabular}
\caption{\textit{Word error rate (WER)[$\%$] on LibriSpeech (LS) settings conducted by Facebook}
}
\end{table}
\subsection{\textit{Tuning the Pseudo-label datasets with WER:}}
Table 2 reports the
comparison results on Test (clean) and Test (other), where, "Filter" indicates the method trained with different filter strategies. No filter and WER (5\% or 10\%) respectively mean that no filtering of predicted pseudo-labels and filtering using a ground-truth label to calculate WER for each sentence. 5\% and 10\% mean that only pseudo-labels with WER less than these two thresholds are selected. LS-100 (Baseline) is the model fine-tuned with 100 hours of labeled data conducted by Facebook. LS-100/LS-860 means that training the model with 100 hours of labeled data and 860 hours of unlabeled data using the IPL semi-supervised learning method.

From the Table 2, we can see, 1) In LS-100 and LS-10 settings, the model trained with pseudo-label data has better performance than the baseline. This verifies the effectiveness of pseudo-label data. 2) The model trained with WER filtering strategy has better performance than it trained with no filtering strategy. This indicates that using high quality pseudo-labels can effectively improve the performance of ASR systems.
\vspace{-1.0em}
\begin{table}[t]
\small
\setlength\tabcolsep{0.5mm}{}
\centering
\label{wer filter}
\renewcommand{\arraystretch}{1}
\begin{tabular}{ccccc}
\toprule
Setting&Model&Filter  & Test(clean)& Test(other)\\
\midrule
LS-100 (Baseline)&\multirow{4}{*}{Wav2vec2.0}&--&6.10&13.30\\
LS-100/LS-860&&No Filter &5.15 &10.86\\
LS-100/LS-860&&WER (5\%) &5.08 &10.62\\
LS-100/LS-860&&WER (10\%) &\textbf{4.98} &\textbf{10.25} \\
\midrule
LS-10 (Baseline)&\multirow{4}{*}{Wav2vec2.0}&--&11.10&17.60\\
LS-10/LS-90&&No Filter  &7.92 &14.10\\
LS-10/LS-90&&WER(5\%) & 7.60 &14.7\\
LS-10/LS-90&&WER(10\%) & \textbf{7.55}&\textbf{13.54}\\
\bottomrule
\end{tabular}
\caption{\textit{Word error rate [$\%$] without filter and within WER filter in IPL on LS test set. Experiments have been done with labeled data at two different scales (LS-100, LS-10)}}

\end{table}
\begin{table}[t]
\centering
\small
\label{tabble::score_filter}
\renewcommand{\arraystretch}{1}
\begin{tabular}{cccc}
\toprule
Setting&Method  & Test(clean)& Test(other)\\
\midrule
LS-100&Wav2Vec2.0\cite{baevski2020wav2vec} &6.10&13.30\\
LS-100/LS-860& PL \cite{kahn2020self} & 7.2 &18.4\\
LS-100/LS-860& MPL \cite{higuchi2021momentum} &8.3 &16.8\\
LS-100/LS-860& PL+CTC \cite{xu2020iterative} &6.0 &10.3\\
LS-100/LS-860& Score Filter &5.10 &10.84\\
LS-100/LS-860& WER Filter(10\%) &\textbf{4.98} &\textbf{10.25}\\
\midrule
LS-10 &Wav2Vec2.0\cite{baevski2020wav2vec} &11.10&17.60\\
LS-10/LS-90 &Score Filter &7.67&13.86\\
LS-10/LS-90&WER Filter &\textbf{7.55}&\textbf{13.54}\\
\bottomrule
\end{tabular}
\caption{\textit{Word error rate [$\%$] on LS test set. Our approach surpass other PL-based method}}
\end{table}
\subsection{\textit{Tuning the Pseudo-label datasets with Score:}}
To demonstrate the robustness of our method, we compare the results with 
some existing methods on pseudo-labels, as shown in Table 3. First of all, our method (either WER filter and Score filter) surpasses other PL methods. This verifies the effectiveness of our method. Both the WER filter method and the Score filter method are better than Wav2vec2.0-based method. And the number of labeled data does not affect the performance of Score filter and WER filter. Finally, the results of Score filter and WER filter are close to each other. Since WER filter is filtered by comparing with the ground-truth label, the pseudo-labels filtered by WER filter has higher confidence, so its result is better than score filter, but the difference is not big, and score filter does not depend on the real label, so it can be put into the real scene easily.
\subsection{\textit{Proof of hypothesis filtering threshold:}}
\begin{figure}[t]
\centering  
\subfigure{
\label{Fig.sub.1}
\includegraphics[width=0.5\linewidth,height = 3cm]{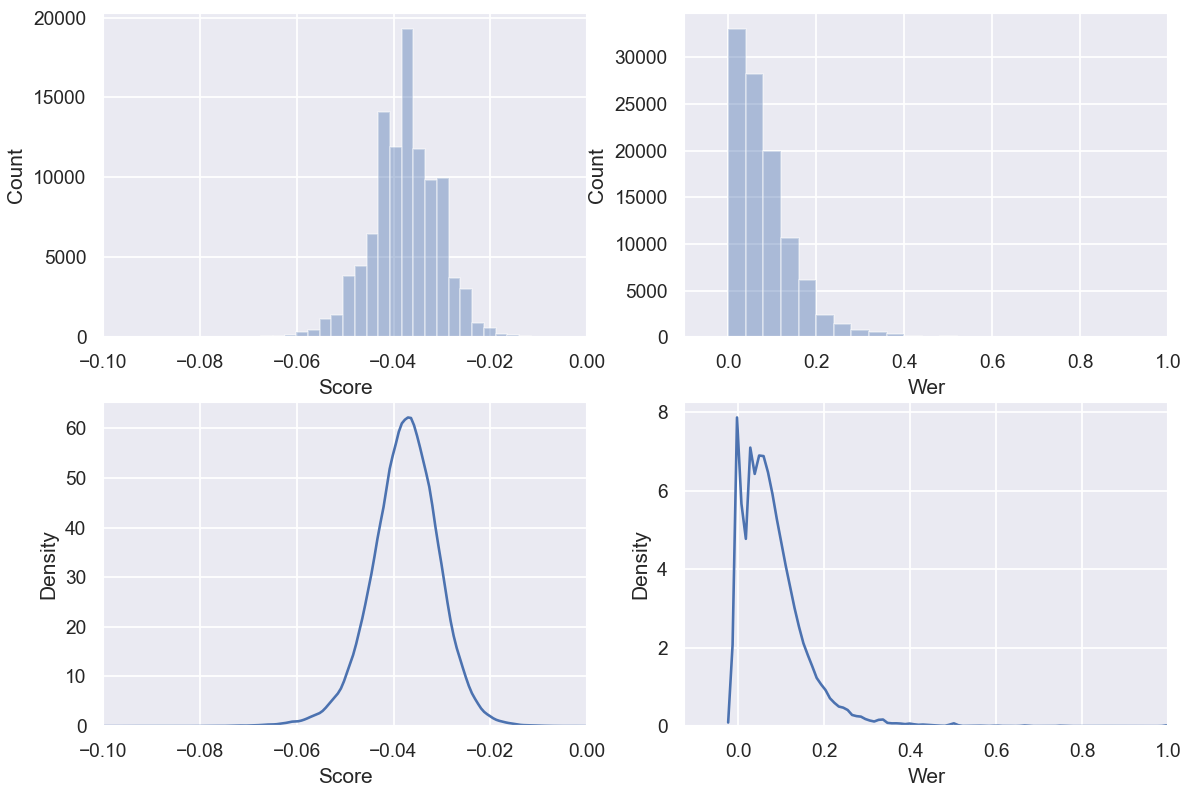}}\subfigure{
\label{Fig.sub.2}
\includegraphics[width=0.5\linewidth,height=3cm]{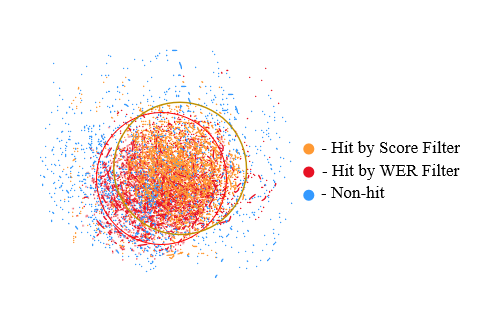}}
\caption{\textbf{Filtered samples distribution after using WER filter and score filter.} In the left part, the top two plots represent the distribution histogram of score and WER. The bottom two plots show the distribution after smoothing through the kernel density function. In the right part, each dot represents a speech data. Orange dots represent the data selected after score filter, red dots show the data selected after WER filter and blue dots indicate the data picked by neither.}
\label{1}

\end{figure}


\vspace{-0.2cm}
\begin{table}[t]
    \centering
    \small
    \label{table 4}
    \setlength\tabcolsep{3mm}{}
    \renewcommand{\arraystretch}{1}
    \begin{tabular}{cccc}
\toprule
Setting&Filtering threshold & test(clean)& test(other)\\
\midrule
\multirow{4}{*}{LS-10/LS-90}&-0.03 &\text{7.87} &\text{14.06}\\

&-0.04 &\text{7.63} &\text{13.81}\\

&-0.05 &\textbf{7.47}&\textbf{12.93}\\

&-0.06 & \text{7.60}&\text{13.73}\\
\bottomrule
\end{tabular}
\caption{\textit{WER[$\%$] under score filter with different filtering threshold on LS test set. Three iterations of training are performed for each filtering threshold, and the one with the best result is selected }}
\end{table}
The pseudo-labels predicted by the model trained on the LS-10/LS-90 setting are the data used in figure \ref{1}. As you can see, there is significant overlap between the red and orange dots; the overlap is estimated to be 72\%. The blue dots are regarded as low-quality pseudo-labels and they are far away from other dots. It means that the pseudo-labels selected by both score filter and WER filter have a high degree of confidence.

According to the score distribution, we set -0.03 as the initial filtering threshold and let the step size equal to 0.01. 
A filtering threshold value is changed from large to small in fixed steps every three iterations. In Table 4, due to the increasing number of training samples, the final performance of the model improves as the threshold decreases. However, since the quality of the training samples decreases, the model performance deteriorates as the threshold decreases from -0.05 to -0.06. Model achieves the best performance when the filtering threshold setting -0.05. 

For the score filter, -0.05 is the optimal filtering threshold. Under this filtering threshold, we count the number of filtered samples which is nearly equal to the number of overlapped samples in figure \ref{1} and we figure out that the filtering threshold we assumed is pretty close to -0.05. This shows that the threshold we assumed can balance the quantity and quality of the filtered samples. Therefore the hypothesis we mentioned in Section 2.3 is approved. In practice, this strategy can be followed as: (1) Predict a small part of labeled data with the teacher model. (2) Filter the pseudo-labels with score filter and WER filter, calculate the overlap rate of their filtered samples, and plot the distribution of score. (3) According to the score distribution, identify a score value that enables the amount of overlapping samples to be almost equal to the number of filtered samples and this is the filtering threshold.
\section{CONCLUSION}
In this paper, we proposed a framework that obtained a better teacher model using self-supervised pre-trained model. By selecting pseudo-labels depend on average probability score, filtered samples have high confidence. To get a balance between quality and quanity of pseudo-labels, we present a strategy to find an optimal filtering threshold. The outcomes of the experiment on the LS evaluation set show that the proposed method outperforms other PL-based approaches. This method can be applied in practice to solve the problem of insufficient annotation data. In the future, we will explore using score filtering method to optimize other ASR systems.

\bibliographystyle{IEEEbib}
\bibliography{icassp}

\end{document}